\documentclass[12pt]{article}

\begin{document}

\input amssym.tex

\title{The physical meaning of the de Sitter invariants}

\author{Ion I. Cot\u aescu \thanks{E-mail:~~cota@physics.uvt.ro}\\
{\it West University of Timi\c soara,}\\{\it V. Parvan Ave. 4,
RO-300223 Timi\c soara}}

\maketitle

\begin{abstract}
We study the Lie  algebras of the covariant representations
transforming the matter fields under the de Sitter isometries. We
point out that the Casimir operators of these representations can be
written in closed forms and we deduce how their eigenvalues depend
on the field's rest energy and spin. For the scalar, vector and
Dirac fields, which have well-defined field equations, we express
these eigenvalues in terms of mass and spin obtaining thus the
principal invariants of the theory of free fields on the de Sitter
spacetime. We show that in the flat limit we recover the
corresponding invariants of the Wigner irreducible representations
of the Poincar\' e group.

Pacs: 04.20.Cv, 04.62.+v, 11.30.-j
\end{abstract}

\vspace*{12mm} Keywords: de Sitter isometries; covariant
representations; Casimir operators; unique spin.

\newpage

\section{Introduction}

In general relativity, apart from the general conservation laws
associated to the relativistic and gauge covariance \cite{FF,FFF},
the isometries play an important role giving rise to the principal
conserved physical quantities \cite{WALD}. The best example is the
de Sitter spacetime \cite{SW,BD} which has $SO(1,4)$ isometries and,
moreover, allows one to analytically solve the equations of the
usual matter fields variously coupled to the de Sitter gravity.
These opportunities encouraged many authors to develop either a de
Sitter quantum mechanics \cite{Gaz} or a quantum field theory
\cite{Wood} in an axiomatic manner, exploiting the high symmetry of
this manifold but avoiding the standard steps of the canonical
theory.

We preferred another strategy, considering the quantum theory of
fields on curved spacetimes based on the canonical quantization.
This is a Lagrangian tetrad-gauge covariant field theory in
non-holonomic orthogonal (local) frames where the fields with spin
half can be correctly defined \cite{SG}. Under such circumstances
the theory of the spacetime symmetries must be completed with the
tools able to control simultaneously the transformations of both the
holonomic and non-holonomic frames. For this reason we introduced
the external symmetry transformations which combine isometries and
gauge transformations such that the tetrad fields remain invariant
under isometries \cite{ES}. We obtained thus the external symmetry
group which is the universal covering group of the isometry one.
Moreover, we pointed out that the matter fields transform under
isometries according to the {\em covariant} representations (CR) of
this group {\em induced} by the linear representations of the gauge
group corresponding to the metric of the  (pseudo) Euclidean model
of the curved manifold. We have shown that the basis-generators of
these CRs  are given by the Carter and McLenaghan formula
\cite{CML}, derived initially for the Dirac field and generalized
then for any CR \cite{ES,EPL}. These operators are important since
they commute with those of the free field equations playing thus the
role of conserved quantities.

Our approach is helpful on the four-dimensional de Sitter  spacetime
where all the usual free field equations can be analytically solved
while the isometries give rise to the rich $so(1,4)$ algebra. Hereby
we selected various sets of commuting operators determining the
quantum modes of the scalar \cite{Cs}, vector \cite{Cv,Max} and
Dirac \cite{CD1,CD2} fields.

We must stress that many of our results differ from those of Refs
\cite{Gaz} where the tetrad-gauge covariance is neglected focusing
only on the usual {\em linear} representations of the isometry
group. In this way one obtains a quantum mechanics on the
five-dimensional Minkowski manifold in which the de Sitter spacetime
is embedded. The advantage therein is that the fields obey
five-dimensional homogeneous equations and transform according to
unitary and irreducible representations (UIR) which have well-known
properties \cite{WW}. Unfortunately, these representations {do not}
play the main role in our gauge-covariant field theory since this is
a four-dimensional theory in the non-holonomic frames of the de
Sitter spacetime where the CRs of the isometry group are {induced}
by the finite-dimensional representations  of the gauge group. For
this reason we believe that the study of the Casimir operators of
our CRs could complete the analyze of the de Sitter invariants
offered by the five-dimensional theory \cite{Gaz}.

The present paper is devoted to this problem. We calculate first the
$SO(1,4)$ generators of the CRs in the de Sitter local chart with
FRW line element and proper time where we choose the diagonal gauge.
These generators are differential operators with spin parts given by
the generators of the finite-dimensional representations of the
$SL(2,{\Bbb C})$ gauge group that induce the CRs. Furthermore, we
write down the Casimir operators of the CRs and analyze their
properties pointing out that their eigenvalues can be expressed in
terms of rest energy and spin when the particles are at rest. A
remarkable identity relating these Casimir operators is derived for
the CRs with unique spin in the sense of the $SL(2,\Bbb C)$ gauge
symmetry. In the flat limit this identity gives just the second
invariant of the  Wigner UIRs of the Poincar\' e group
\cite{Wigner,W}. These results combined with the definitions of the
masses given by the field equations enable us to write down the de
Sitter invariants of the basic fields in terms of mass and spin, as
in special relativity, but with supplemental terms which vanishes in
the flat limit.

We start in the second section with a short review of our theory of
external symmetry following to briefly present in the next one the
Poincar\' e invariants determined by the mass and spin according to
the Wigner theory of the UIRs of this group \cite{Wigner}. The
principal new results are presented in section 4 where we calculate
the $SO(1,4)$ generators of the CRs  deriving the corresponding
Casimir operators in closed forms and finding their eigenvalues for
the particles at rest. Moreover, we establish the mentioned identity
among these operators and the spin ones that holds in many cases
when the spin is unique. In the next section we deduce the de Sitter
invariants of the scalar, vector and Dirac fields minimally coupled
to the de Sitter gravity. Our conclusions are presented in the last
section. The Appendix is devoted to the classical conserved
quantities which can be compared with the quantum ones.

\section{External symmetry}

The theory of the matter fields with spin on the pseudo-Riemannian
spacetime $(M,g)$ can be formulated considering simultaneously both
the holonomic and non-holonomic frames. The holonomic  frames are
local charts of coordinates $x^{\mu}$, labeled by natural indices,
$\mu, \nu,...=0,1,2,3$. In a given tetrad-gauge, the tetrad fields
$e_{\hat\mu}$  and  $\hat e^{\hat\mu}$, which define the
non-holonomic orthogonal frames and the corresponding coframes, are
labeled by local indices, $\hat\mu, \hat\nu,...=0,1,2,3$. These
fields have the usual duality, $\hat e^{\hat\mu}_{\alpha}\,
e_{\hat\nu}^{\alpha}=\delta^{\hat\mu}_{\hat\nu}$, $ \hat
e^{\hat\mu}_{\alpha}\, e_{\hat\mu}^{\beta}=\delta^{\beta}_{\alpha}$,
and orthonormalization,  $e_{\hat\mu}\cdot e_{\hat\nu}=\eta_{\hat\mu
\hat\nu}$, $\hat e^{\hat\mu}\cdot \hat e^{\hat\nu}=\eta^{\hat\mu
\hat\nu}$,  properties. The metric tensor $g_{\mu
\nu}=\eta_{\hat\alpha\hat\beta}\hat e^{\hat\alpha}_{\mu}\hat
e^{\hat\beta}_{\nu}$ raises or lowers the natural indices while for
the local indices  we have to use the flat metric
$\eta=$diag$(1,-1,-1,-1)$ of the Minkowski spacetime $(M_0,\eta)$
which is the pseudo-Euclidean model of $(M,g)$.

The metric $\eta$ remains invariant under the transformations of the
group $O(1,3)$ which includes as a subgroup the Lorentz group,
$L_{+}^{\uparrow}$, whose universal covering group is $SL(2,\Bbb
C)$. In the usual covariant parametrization, with the real
parameters, $\omega^{\hat\alpha
\hat\beta}=-\omega^{\hat\beta\hat\alpha}$, the transformations
$A(\omega)=\exp(-\frac{i}{2}\omega^{\hat\alpha\hat\beta}
S_{\hat\alpha\hat\beta}) \in SL(2,\Bbb C)$ depend on the covariant
basis-generators of the $sl(2,\Bbb C)$ Lie algebra, $S_{\hat\alpha
\hat\beta}$, which are the principal spin operators generating all
the spin terms of other operators. This parametrization offers, in
addition, the advantage of a simple expansion of the matrix elements
in local bases,
$\Lambda^{\hat\mu\,\cdot}_{\cdot\,\hat\nu}[A(\omega)]=
\delta^{\hat\mu}_{\hat\nu}
+\omega^{\hat\mu\,\cdot}_{\cdot\,\hat\nu}+\cdots$, of the
transformations $\Lambda[A(\omega)]\in L_{+}^{\uparrow}$ associated
to $A(\omega)$ through the canonical homomorphism \cite{W}.

Assuming now that $(M,g)$ is orientable and time-orientable we can
restrict ourselves to consider $G(\eta)=L^{\uparrow}_{+}$  as the
gauge group of the Minkowski metric $\eta$ \cite{WALD}. This is the
structure group of the principal fiber bundle whose basis is $M$
while the group ${\rm Spin}(\eta)=SL(2,\Bbb C)$  represents the
structure group of the spin fiber bundle \cite{WALD,SG} which is
well-defined when some special conditions are fulfilled as in the
case of the de Sitter manifold that is globally hyperbolic
\cite{Geroch}. The matter fields, $\psi_{(\rho)}:\, M\to {\cal
V}_{(\rho)}$, are locally defined over $M$ with values in the vector
spaces ${\cal V}_{(\rho)}$ carrying finite-dimensional
representations $\rho$ of the group $SL(2,\Bbb C)$ which, in
general, are reducible as direct sums of irreducible ones,
$(j_+,j_-)$ \cite{W}. These representations are non-unitary and,
consequently, in order to assure the global unitary properties of
the field theory one must use invariant hermitian forms called often
relativistic scalar products \footnote{We proposed a general
solution of this problem in {\tt arXiv:math-ph/9904029} but this
paper was never published because of its 'horrific' notations.}
which play a similar role in special or general relativity.

The choice of the representation $\rho$ determines the form of the
covariant derivatives of the field $\psi_{(\rho)}$ in local frames,
\begin{equation}\label{der}
D_{\hat\alpha}^{(\rho)}= e_{\hat\alpha}^{\mu}D_{\mu}^{(\rho)}=
\hat\partial_{\hat\alpha}+\frac{i}{2}\, \rho(S^{\hat\beta\, \cdot} _{\cdot\,
\hat\gamma})\,\hat\Gamma^{\hat\gamma}_{\hat\alpha \hat\beta}\,,
\end{equation}
that depend, in addition,  on the local derivatives
$\hat\partial_{\hat\alpha}=e^{\mu}_{\hat\alpha}\partial_{\mu}$ and
the connection coefficients in local frames,
$\hat\Gamma^{\hat\sigma}_{\hat\mu \hat\nu}=e_{\hat\mu}^{\alpha}
e_{\hat\nu}^{\beta}(\hat e_{\gamma}^{\hat\sigma}
\Gamma^{\gamma}_{\alpha \beta} -\hat e^{\hat\sigma}_{\beta,
\alpha})$, which assure the covariance of the whole theory under
tetrad-gauge transformations produced by the automorphisms $A\in
SL(2,{\Bbb C})$ of the spin fiber bundle. This is the general
framework of the theories involving fields with half integer spin
which can not be formulated in natural frames.

A special difficulty of the field theory in local frames arises from
the fact that the tetrad fields transform under isometries in a
non-covariant manner because of their natural indices. For this
reason we proposed the theory of external symmetry in which each
isometry transformation is coupled to a gauge one able to correct
the position of the local frames such that the whole transformation
should preserve not only the metric but the tetrad-gauge too
\cite{ES}. Thus, for any isometry transformation $x\to
x'=\phi_{\xi}(x)=x+\xi^a k_a(x) +...$, depending on the parameters
$\xi^a$ ($a,b,...=1,2...N$) of the isometry group $I(M)$, one must
perform the gauge transformation $A_{\xi}$ defined as
\begin{equation}\label{Axx}
\Lambda^{\hat\alpha\,\cdot}_{\cdot\,\hat\beta}[A_{\xi}(x)]= \hat
e_{\mu}^{\hat\alpha}[\phi_{\xi}(x)]\frac{\partial
\phi^{\mu}_{\xi}(x)} {\partial x^{\nu}}\,e^{\nu}_{\hat\beta}(x)\,,
\end{equation}
with the supplementary condition $A_{\xi=0}(x)=1\in SL(2,\Bbb C)$.  Then the
transformation laws of our fields are
\begin{equation}\label{es}
(A_{\xi},\phi_{\xi}):\qquad
\begin{array}{rlrcl}
e(x)&\to&e'(x')&=&e[\phi_{\xi}(x)]\,,\\
\hat e(x)&\to&\hat e'(x')&=&\hat e[\phi_{\xi}(x)]\,,\\
\psi_{(\rho)}(x)&\to&\psi_{(\rho)}'(x')&=&\rho[A_{\xi}(x)]\psi_{(\rho)}(x)\,.
\end{array}
\qquad
\end{equation}
We have shown that the pairs $(A_{\xi},\phi_{\xi})$ constitute a
well-defined Lie group we called the external symmetry group of
$(M,g)$, denoted by $S(M)$, pointing out that this is just the
universal covering group of $I(M)$ \cite{ES}. For small values of
$\xi^{a}$, the $SL(2,\Bbb C)$ parameters of $A_{\xi}(x)\equiv
A[\omega_{\xi}(x)]$ can be expanded as
$\omega^{\hat\alpha\hat\beta}_{\xi}(x)=
\xi^{a}\Omega^{\hat\alpha\hat\beta}_{a}(x)+\cdots$, in terms of the
functions
\begin{equation}\label{Om}
\Omega^{\hat\alpha\hat\beta}_{a}\equiv {\frac{\partial
\omega^{\hat\alpha\hat\beta}_{\xi}} {\partial\xi^a}}_{|\xi=0}
=\left( \hat e^{\hat\alpha}_{\mu}\,k_{a,\nu}^{\mu} +\hat
e^{\hat\alpha}_{\nu,\mu}
k_{a}^{\mu}\right)e^{\nu}_{\hat\lambda}\eta^{\hat\lambda\hat\beta}
\end{equation}
which depend on the Killing vectors
$k_a=\partial_{\xi_a}\phi_{\xi}|_{\xi=0}$ associated to the
parameters $\xi^a$. We note that the functions (\ref{Om}) are
skew-symmetric,
$\Omega^{\hat\alpha\hat\beta}_{a}=-\Omega^{\hat\beta\hat\alpha}_{a}$,
only when $k_a$ are Killing vectors \cite{ES}.

The last of equations (\ref{es})  defines the operator-valued
representations $T^{(\rho)} \,:\, (A_{\xi},\phi_{\xi})\to
T_{\xi}^{(\rho)}$ of the group $S(M)$ which are called the {\em
covariant} representations (CR) {\em induced} by the
finite-dimensional representations $\rho$ of the group $SL(2,\Bbb
C)$. The covariant transformations,
\begin{equation}
(T_{\xi}^{(\rho)}\psi_{(\rho)})[\phi_{\xi}(x)]=\rho[A_{\xi}(x)]\psi_{(\rho)}(x)\,,
\end{equation}
leave the field equation invariant since their basis-generators
\cite{ES},
\begin{equation}\label{Sx}
X_{a}^{(\rho)}=i{\partial_{\xi^a} T_{\xi}^{(\rho)}}_{|\xi=0}=-i
k_a^{\mu}\partial_{\mu} +\frac{1}{2}\,\Omega^{\hat\alpha\hat\beta}_{a}
\rho(S_{\hat\alpha\hat\beta})\,,
\end{equation}
commute with the operator of the field equation and satisfy the
commutation rules $[X_{a}^{(\rho)},
X_{b}^{(\rho)}]=ic_{abc}X_{c}^{(\rho)}$ determined by the structure
constants, $c_{abc}$, of the algebras $s(M)\sim i(M)$. In other
words, the operators (\ref{Sx}) are the basis-generators of a CR of
the $s(M)$ algebra {\em induced} by the representation $\rho$ of the
$sl(2,{\Bbb C})$ algebra.

We note that these generators are proportional to the  Kosmann's Lie
derivatives \cite{Kos} associated to the Killing vectors $k_a$. They
can be put in  covariant form either in non-holonomic frames
\cite{ES} or even in holonomic ones \cite{EPL}, generalizing thus
the formula given by Carter and McLenaghan for the Dirac field
\cite{CML}.

A specific feature of the CRs is that their generators have, in
general, point-dependent spin terms which do not commute with the
orbital parts. However, there are tetrad-gauges in which at least
the generators of a subgroup  $G \subset I(M)$ may have
point-independent spin terms commuting with the orbital parts. Then
we say that the restriction to $G$ of the CR $T^{(\rho)}$ is {\em
manifest} covariant \cite{ES}. Obviously, if $G=I(M)$ then the whole
representation $T^{(\rho)}$ is manifest covariant.

\section{The Poincar\' e invariants}

In this paper we study the physical meaning of the de Sitter
invariants of the fields which transforms according to the CRs
$T^{(\rho)}$. In the flat limit these invariants must coincide with
the well-known  Poincar\' e invariants which are completely
determined by the mass and spin of the matter field. For this reason
it deserves to briefly review the properties of these invariants
focusing only on the massive case.

On the Minkowski flat spacetime $(M_0,\eta)$ the fields
$\psi_{(\rho)}$ transform under isometries according to manifest
covariant representations in {\em inertial} (local) frames defined
by $e^{\mu}_{\nu}=\hat e^{\mu}_{\nu}=\delta^{\mu}_{\nu}$. The
isometries are just the transformations $x\to x'=\Lambda[A(\omega)]x
-a$ of the Poincar\' e group $I(M_0)= {\cal P}_{+}^{\uparrow} =
T(4)\circledS L_{+}^{\uparrow}$ \cite{W} whose universal covering
group is $S(M_0)= \tilde{\cal P}^{\uparrow}_{+}=T(4)\circledS
SL(2,\Bbb C)$. Both these groups are semidirect products where the
translations form the {\em normal} Abelian subgroup $T(4)$.

The manifest covariant representations,  $ T^{(\rho)} \,:\,
(A(\omega), a)\to  T_{\omega,a}^{(\rho)}$, of the group $S(M_0)$
have the transformation rules
\begin{equation}\label{TPoin}
(T_{\omega,a}^{(\rho)}\psi_{(\rho)})(x)
=\rho[A(\omega)]\psi_{(\rho)}\{\Lambda[A(\omega)]^{-1}(x+a)\},
\end{equation}
and the well-known basis-generators of the $s(M_0)$ algebra,
\begin{eqnarray}
\hat P_{\mu}& \equiv &\hat X_{(\mu)}^{(\rho)}=i\partial_{\mu}\,, \label{XMin1}\\
\hat J_{\mu\nu}^{(\rho)}&\equiv&\hat X_{(\mu\nu)}^{(\rho)}
=(\eta_{\mu\alpha}x^{\alpha}
\partial_{\nu}- \eta_{\nu\alpha}x^{\alpha}\partial_{\mu})+
S_{\mu\nu}^{(\rho)}\,,\label{XMin2}
\end{eqnarray}
which have point-independent spin parts denoted from now by
$S_{\hat\mu\hat\nu}^{(\rho)}$ instead of $\rho(S_{\hat\mu\hat\nu})$.
Here it is convenient to separate the energy operator, $\hat H=\hat
P_0$, and to write the $sl(2,{\Bbb C})$ generators as
\begin{eqnarray}
\hat J_i^{(\rho)}&=&\frac{1}{2}\,\varepsilon_{ijk}\hat
J^{(\rho)}_{jk}=
-i\varepsilon_{ijk}x^j\partial_k+S_i^{(\rho)}\,,\quad
S_i^{(\rho)}=\frac{1}{2}\,
\varepsilon_{ijk}S_{jk}^{(\rho)}\,,\\
\hat K_i^{(\rho)}&=&\hat J_{0i}^{(\rho)}=i (x^i
\partial_t+t\partial_i)+S_{0i}^{(\rho)}\,,\qquad i,j,k...=1,2,3\,,
\end{eqnarray}
denoting $\vec{S}^2=S_iS_i$ and $\vec{S}_0^2=S_{0i}S_{0i}$. Thus we
lay out the standard basis of the $s(M_0)$ algebra, $\{\hat H, \hat
P_i,\hat J^{(\rho)}_i,\hat K^{(\rho)}_i\}$.

The invariants of the manifest covariant fields are  the eigenvalues
of the Casimir operators of the representations $T^{(\rho)}$ that
read
\begin{equation}\label{CM}
\hat {\cal C}_1=\hat P_{\mu}\hat P^{\mu}\,,\quad \hat {\cal
C}_2^{(\rho)}=-\eta_{\mu\nu}\hat W^{(\rho)\,\mu}\hat
W^{(\rho)\,\nu}\,,
\end{equation}
where the Pauli-Lubanski operator \cite{W},
\begin{equation}\label{PaLu}
\hat
W^{(\rho)\,\mu}=-\frac{1}{2}\,\varepsilon^{\mu\nu\alpha\beta}\hat
P_{\nu} \hat J_{\alpha\beta}^{(\rho)}\,,
\end{equation}
has the components
\begin{equation}
\hat W^{(\rho)}_0={\hat J}^{(\rho)}_i{\hat P}_i={S}^{(\rho)}_i{\hat
P}_i\,,\quad \hat W^{(\rho)}_i=\hat H\,{\hat J}_i^{(\rho)}+
\varepsilon_{ijk}{\hat K}^{(\rho)}_j {\hat P}_k\,,
\end{equation}
resulting from equations (\ref{XMin1}) and (\ref{XMin2}) where we
take $\varepsilon^{0123}=-\varepsilon_{0123}=-1$.

The first invariant (\ref{CM}a) deals with the mass condition, $\hat
P^2\psi_{(\rho)}=m^2\psi_{(\rho)}$, fixing the orbit in the momentum
spaces on which the Fourier transform, $\tilde \psi_{(\rho)}$, of
the covariant field $\psi_{(\rho)}$ is defined. In general, the
field  $\tilde \psi_{(\rho)}$ transforms under isometries according
to a reducible CR, $\tilde T^{(\rho)}$, which is equivalent with a
direct sum of Wigner UIRs, $(\pm m,s)$, induced by the subgroup
$T(4) \circledS SU(2)\subset SL(2,{\Bbb C})$ \cite{Wigner,W}.
Consequently, each subspace ${\cal V}_s\subset {\cal V}_{(\rho)}$ of
spin $s$ carries its own UIRs $(\pm m,s)$ for which the second
Casimir operator yields the eigenvalue \cite{W}
\begin{equation}\label{ident}
\hat {\cal C}_2^{(s)}\sim m^2 s(s+1)\,.
\end{equation}
These invariants can be easily calculated in the {\em rest} frames
where $\hat P_i\sim 0$ so that $\hat W^{(\rho)}_0\sim 0$ and $\hat
W^{(\rho)}_i\sim \hat H S^{(\rho)}_i$ while the mass condition gives
the rest energy $\hat H\sim E_0 =\pm\,m$. As a matter of fact, the
Wigner UIRs of the massive particles (and antiparticles) are
determined by $E_0$ and $s$.

The advantage of this approach is that the UIRs $(\pm m,s)$ are
unitary with respect to a scalar product in momentum representation
which corresponds to the relativistic scalar products of the
covariant fields $\psi_{(\rho)}$. Moreover, they point out the mass
and spin as being the principal invariants of the theory of free
fields and help one to keep under control the spin content of the
covariant representations. When $\tilde T^{(s)}$ is equivalent only
with the UIRs $(\pm m,s)$ one says that the covariant field has the
unique spin $s$.

\section{The de Sitter invariants}

The Wigner theory in momentum representation works in any (pseudo)
Euclidean manifold whose isometry group is the semidirect product
between an orthogonal subgroup and a normal Abelian one
\cite{Wigner}. Since the isometry group of the Sitter manifold does
not have this property, we must study the CRs  of this group in the
configuration space, following step by step the method presented in
the second section.

\subsection{Killing vectors}

Let us consider $(M,g)$ be the de Sitter spacetime defined as the
hyperboloid of radius $1/\omega$ \footnote{We denote by $\omega $
the Hubble de Sitter constant since  $H$ is reserved for the
Hamiltonian operator} in the five-dimensional flat spacetime
$(M^5,\eta^5)$ of coordinates $z^A$  (labeled by the indices
$A,\,B,...= 0,1,2,3,4$) and metric $\eta^5={\rm
diag}(1,-1,-1,-1,-1)$ \cite{SW}. The local charts of coordinates
$\{x\}$  can be easily introduced on $(M,g)$ giving the set of
functions $z^A(x)$ which solve the hyperboloid equation,
\begin{equation}\label{hip}
\eta^5_{AB}z^A(x) z^B(x)=-\frac{1}{\omega^2}\,.
\end{equation}

In this manner $(M,g)$ is defined  as a homogeneous space of the
pseudo-orthogonal group $SO(1,4)$ which is in the same time the
gauge group of the metric $\eta^{5}$ and the isometry group, $I(M)$,
of the de Sitter spacetime. The group of the external symmetry,
$S(M)={\rm Spin}(\eta^5)=Sp(2,2)$, has the Lie algebra
$s(M)=sp(2,2)\sim so(1,4)$ for which  we use the covariant real
parameters $\xi^{AB}=-\xi^{BA}$. Then,  the orbital basis-generators
of the natural representation of the $s(M)$ algebra (carried by the
space of the scalar functions over $M^{5}$) have the standard form
\begin{equation}\label{LAB5}
 L_{AB}^{5}=i\left[\eta_{AC}^5 z^{C}\partial_{B}-
 \eta_{BC}^5 z^{C}
\partial_{A}\right]=-iK_{(AB)}^C\partial_C
\end{equation}
which allows us to derive the corresponding Killing vectors of
$(M,g)$, $k_{(AB)}$, using the identities $k_{(AB)
\mu}dx^{\mu}=K_{(AB) C}dz^C$.

We assume that $(M,g)$ is equipped with the local chart
$\{t,\vec{x}\}$ of Cartesian coordinates defined by the functions
\begin{eqnarray}
z^0(x)&=&\chi(x)\, e^{\omega t}\,,\quad
\chi(x)=\frac{1}{2\omega}(1+\omega^2 \vec{x}^2-e^{-2\omega t})\,,\label{z0}\\
z^4(x)&=&-\chi(x)\, e^{\omega t}+\frac{1}{\omega}\,e^{\omega
t}\,,\quad z^i(x)=e^{\omega t} x^i \,,
\end{eqnarray}
giving rise to the FRW line element
\begin{equation}\label{FRW}
ds^2=\eta^5_{AB}dz^A dz^B=g_{\mu\nu}dx^{\mu}dx^{\nu}=dt^2-e^{2\omega
t}(d\vec{x}\cdot d\vec{x})\,.
\end{equation}
In this chart the Killing vectors of the de Sitter symmetry have the
components:
\begin{eqnarray}
&&k^0_{(0i)}=k^0_{(4i)}=x^i\,,\qquad k^j_{(0i)}=k^j_{(4i)}-
\frac{1}{\omega}\,\delta^j_i=\omega x^ix^j-\delta^j_i\chi\,,\label{chichi1}\\
&&k^0_{(ij)}=0\,,\quad k^l_{(ij)}=\delta^l_jx^i-\delta^l_i
x^j\,;\qquad k^0_{(04)}=-\frac{1}{\omega}\,,\quad
k^i_{(04)}=x^i\,.\label{chichi2}
\end{eqnarray}
The form of the basis-generators of our CRs depends on these
components and on the choice of the tetrad-gauge. The simplest gauge
is the diagonal one in which the non-vanishing components of the
tetrad fields are
\begin{equation}\label{gauge}
\hat e^0_0=e^0_0=1\,, \quad \hat e^i_j=e^{\omega
t}\delta^i_j\,,\quad e^i_j=e^{-\omega t}\delta^i_j\,.
\end{equation}

\subsection{Generators of covariant representations}

According to the general theory, the generators of the CRs
$T^{(\rho)}$ of the group $S(M)=Sp(2,2)$, induced by the
representations $\rho$ of the $SL(2,\Bbb C)$  group, constitute CRs
of the $sp(2,2)$ algebra induced by the representations $\rho$ of
the $sl(2,{\Bbb C})$ algebra. Therefore, their commutation relations
are determined by the structure constant of the group $Sp(2,2)$ and
the principal invariants are the Casimir operators of the CRs which
can be derived as those of the algebras $sp(2,2)\sim so(1,4)$.

In the covariant parametrization of the $sp(2,2)$ algebra adopted
here, the generators $X_{(AB)}^{(\rho)}$ corresponding to the
Killing vectors $k_{(AB)}$  result from  equation (\ref{Sx}) and the
functions (\ref{Om}) with the new labels $a\to (AB)$. Using then the
Killing vectors (\ref{chichi1}) and (\ref{chichi2}) and the
tetrad-gauge (\ref{gauge}) of the chart $\{t,\vec{x}\}$, after a
little calculation, we find first the $sl(2,{\Bbb C})$ generators.
These are the total angular momentum,
\begin{equation}\label{Ji}
J^{(\rho)}_i \equiv  \frac{1}{2}\,\varepsilon_{ijk}
X_{(jk)}^{(\rho)}=-i\varepsilon_{ijk}x^j\partial_k+S_i^{(\rho)}\,,
\end{equation}
and the generators of the Lorentz boosts
\begin{equation}\label{Ki}
K^{(\rho)}_i \equiv  X_{(0i)}^{(\rho)}=i x^i\partial_t+
i\chi(x)\partial_i -i\omega x^i\, {x}^j {\partial}_j+e^{-\omega
t}S_{0i}^{(\rho)}+\omega S_{ij}^{(\rho)}x^j\,,
\end{equation}
where $\chi$ is defined by equation (\ref{z0}b).  In addition, there
are three generators,
\begin{equation}\label{Ai}
R^{(\rho)}_i  \equiv
X_{(i4)}^{(\rho)}=-K^{(\rho)}_i+\frac{1}{\omega}\, i\partial_i\,,
\end{equation}
which play the role of a Runge-Lenz vector, in the sense that $\{
J_i,R_i\}$ generate a $so(4)$ subalgebra. The energy (or
Hamiltonian) operator \cite{CD1},
\begin{equation}\label{Ham}
H \equiv  \omega X_{(04)}^{(\rho)}=i\partial_t-i\omega {x}^i
{\partial}_i\,,
\end{equation}
is given by the Killing vector $k_{(04)}$ which is time-like only
for $\omega |\vec{x}|e^{\omega t}\leq 1$. Fortunately, this
condition is accomplished everywhere inside the light-cone of an
observer at rest in $\vec{x}=0$. Therefore,  despite of some doubts
appeared in literature \cite{Witt}, the operator $H$ is correctly
defined.

The generators introduced above form the basis $\{
H,J^{(\rho)}_i,K^{(\rho)}_i,R^{(\rho)}_i\}$ of the covariant
representation of the $sp(2,2)$ algebra with the following
commutation rules:
\begin{eqnarray} &\left[ J_i^{(\rho)},
J_j^{(\rho)}\right]=i\varepsilon_{ijk} J_k^{(\rho)}\,,\quad&\left[
J_i^{(\rho)},
R_j^{(\rho)}\right]=i\varepsilon_{ijk} R_k^{(\rho)}\,,\label{ALG1}\\
&\left[ J_i^{(\rho)}, K_j^{(\rho)}\right]=i\varepsilon_{ijk}
K_k^{(\rho)}\,,\quad&\left[ R_i^{(\rho)},
R_j^{(\rho)}\right]=i\varepsilon_{ijk}
J_k^{(\rho)}\,,\label{ALG2}\\
&\left[ K_i^{(\rho)}, K_j^{(\rho)}\right]=-i\varepsilon_{ijk}
J_k^{(\rho)}\,,\quad&\left[ R_i^{(\rho)},
K_j^{(\rho)}\right]=\frac{i}{\omega}\,\delta_{ij}H\,,\label{ALG3}
\end{eqnarray}
and
\begin{equation}\label{HHKR}
\left[ H, J_i^{(\rho)}\right]=0\,,\quad \left[ H,
K_i^{(\rho)}\right]=i\omega R_i^{(\rho)}\,,\quad\left[ H,
R_i^{(\rho)}\right]=i\omega K_i^{(\rho)}\,.
\end{equation}
In some applications it is useful to replace the operators
$\vec{K}^{(\rho)}$ and $\vec{R}^{(\rho)}$ by the momentum operator
$\vec{P}$ and its dual, $\vec{Q}^{(\rho)}$, whose components are
defined as
\begin{equation}\label{Pi}
P_i= \omega(R^{(\rho)}_i+K^{(\rho)}_i)=i\partial_i\,, \quad
Q_i^{(\rho)}= \omega(R^{(\rho)}_i-K^{(\rho)}_i)\,.
\end{equation}
We obtain thus the basis $\{H, P_i,Q^{(\rho)}_i,J^{(\rho)}_i\}$ with
the new commutators
\begin{eqnarray}
&&\left[ H, P_i \right]=i\omega P_i\,,\quad \hspace*{8.5 mm} \left[
H, Q^{(\rho)}_i
\right]=-i\omega Q^{(\rho)}_i\,,\label{HPQ}\\
&&\left[J^{(\rho)}_i , P_j \right]=i\varepsilon_{ijk} P_k\,,\quad
\left[ J^{(\rho)}_i, Q^{(\rho)}_j
\right]=i\varepsilon_{ijk} Q^{(\rho)}_k\,,\label{PJP}\\
&&\left[Q_i^{(\rho)}, P_j \right]=2 i \omega \delta_{ij} H + 2 i
\omega^2 \varepsilon_{ijk} J^{(\rho)}_k \,,\\
&&\left[Q_i^{(\rho)}, Q_j^{(\rho)} \right]=[P_i,P_j]=0\,.
\end{eqnarray}
Another basis is of the Poincar\' e type being formed by  $\{H,
P_i,J^{(\rho)}_i,K^{(\rho)}_i\}$. This has the commutation rules
given by equations (\ref{ALG1}a), (\ref{ALG2}a), (\ref{ALG3}a),
(\ref{HPQ}a), (\ref{PJP}a) and
$\left[P_i,K^{(\rho)}_j\right]=i\delta_{ij}
H-i\omega\varepsilon_{ijk}J^{(\rho)}_k$, while the commutator
(\ref{HHKR}b) has to be rewritten as $\left[ H,
K_i^{(\rho)}\right]=i P_i-i\omega K_i^{(\rho)}$.

The last two bases bring together the conserved energy (\ref{Ham})
and momentum (\ref{Pi}a) which are the only genuine orbital
operators, independent on $\rho$. What is specific for the de Sitter
symmetry is that these operators can not be put simultaneously in
diagonal form since, according to equation (\ref{HPQ}a), they do not
commute to each other. Therefore, there are no mass-shells
\cite{CD1}.

\subsection{Casimir operators}

The first invariant of the CR $T^{(\rho)}$ is the quadratic Casimir
operator
\begin{eqnarray}
{\cal C}^{(\rho)}_1&=&-\,\omega^2
\frac{1}{2}\,X_{(AB)}^{(\rho)}X^{(\rho)\,(AB)}\label{Q1}\\
&=&H^2-\omega^2({\vec{J}^{(\rho)}}\cdot
{\vec{J}^{(\rho)}}+{\vec{R}^{(\rho)}}\cdot
{\vec{R}^{(\rho)}}-{\vec{K}^{(\rho)}}\cdot{\vec{K}^{(\rho)}})\label{C14}\\
&=&H^2+3 i \omega H-{\vec{Q}^{(\rho)}}\cdot{\vec{P}}
-\omega^2{\vec{J}^{(\rho)}}\cdot{\vec{J}^{(\rho)}}\,.\label{C13}
\end{eqnarray}
which can be calculated according to equations
(\ref{Ji})-(\ref{Ham}) and (\ref{Pi}). After a few manipulation we
obtain its definitive expression
\begin{equation}\label{CES}
{\cal C}_1^{(\rho)}={\cal E}_{KG}+2i\omega e^{-\omega
t}S_{0i}^{(\rho)}\partial_i-\omega^2({\vec{S}}^{(\rho)})^2\,,
\end{equation}
depending on the Klein-Gordon operator of the scalar field,
\begin{equation}\label{EKG}
{\cal E}_{KG}= -\partial_t^2-3\,\omega\partial_t+e^{-2\omega
t}\Delta\,, \quad\Delta={\vec{\partial}\,}^2\,.
\end{equation}

The second Casimir operator,
\begin{equation}\label{Q2}
{\cal C}^{(\rho)}_2=-\eta^5_{AB}W^{(\rho)\, A}W^{(\rho)\,B}\,,
\end{equation}
is written with the help of the five-dimensional vector-operator
$W^{(\rho)}$ whose components read \cite{Gaz}
\begin{equation}\label{WW}
W^{(\rho)\, A}=\frac{1}{8}\,\omega\, \varepsilon^{ABCDE}
X_{(BC)}^{(\rho)}X_{(DE)}^{(\rho)}\,,
\end{equation}
where $\varepsilon^{01234}=1$ and the factor $\omega$  assures the
correct flat limit.  After a little calculation we obtain the
concrete form of these components,
\begin{eqnarray}
W^{(\rho)}_0&=&\,\omega\, \vec{J}^{(\rho)}\cdot\vec{R}^{(\rho)}
\,,\label{WW0}\\
{W}^{(\rho)}_i&=&H\,{J}_i^{(\rho)}+\omega\,
\varepsilon_{ijk}{K}^{(\rho)}_j {R}_k^{(\rho)}\,,\\
W^{(\rho)}_4&=&\,-\omega\,
\vec{J}^{(\rho)}\cdot\vec{K}^{(\rho)}\,,\label{WW5}
\end{eqnarray}
which indicate that $W^{(\rho)}$  plays an important role in
theories with spin, similar to that of the Pauli-Lubanski operator
(\ref{PaLu}) of the Poincar\' e symmetry. For example, the helicity
operator is now $W^{(\rho)}_0-W^{(\rho)}_4={S}^{(\rho)}_i {P}_i$.

Replacing then the components (\ref{WW0})-(\ref{WW5}) in equation
(\ref{Q2}) we are faced with a complicated calculation but which can
be performed using algebraic codes under Maple. Thus we obtain the
closed form of the second Casimir operator,
\begin{eqnarray}
{\cal C}_2^{(\rho)}&=& -(\vec{S}^{(\rho)})^2(\partial_t^2+3\,\omega
\partial_t + 2\omega^2) +
2 e^{-\omega t}(iS_{0k}^{(\rho)}-\varepsilon_{ijk}S_i^{(\rho)}
S_{0j}^{(\rho)})\partial_k\partial_t\nonumber\\
&&-e^{-2\omega t}\left[(\vec{S_0}^{(\rho)})^2\Delta
-(S_i^{(\rho)}S_j^{(\rho)}+S_{0i}^{(\rho)}S_{0j}^{(\rho)})
\partial_i\partial_j\right]\nonumber\\
&&+2i\omega e^{-\omega t}( S_i^{(\rho)} S_k^{(\rho)}
S_{0i}^{(\rho)}+S_{0k}^{(\rho)})\partial_k\,,\label{Q22}
\end{eqnarray}
which represents a step forward to a complete theory of the de
Sitter invariants of the covariant fields.

Hence, we derived the general expressions of both the Casimir
operators of the CRs $T^{(\rho)}$  of the group $Sp(2,2)$ induced by
the finite-dimensional representations $\rho$ of the $SL(2,\Bbb C)$
group. In general, these operators are complicated and can not be
related to each other in an arbitrary frame and for any
representation $\rho$. On the other hand, here we do not have a
theory of reducibility in momentum representation similar to the
Wigner one. Therefore,  we must restrict ourselves to study these
invariants in the configuration space or at most by using momentum
expansions in which only $P_i$ are diagonal while $H$ does not do
that.

It is interesting to look for the invariants of the particles at
rest in the chart $\{t,\vec{x}\}$. These have the vanishing momentum
($P_i\sim 0$) so that $H$ acts as $i\partial_t$ and, therefore, it
can be put in diagonal form its eigenvalues being just the rest
energies, $E_0$. Then, for each subspace  ${\cal V}_s \subset{\cal
V}_{(\rho)}$ of given spin, $s$, we obtain the eigenvalues of the
first Casimir operator,
\begin{equation}\label{rest}
{\cal C}_1^{(\rho)}\sim E_0^2+3i\omega E_0 -\omega^2 s(s+1)\,,
\end{equation}
using equations (\ref{CES}) and (\ref{EKG}) while those of the
second Casimir operator,
\begin{equation}\label{rest2}
{\cal C}_2^{(\rho)}\sim s(s+1)(E_0^2+3i\omega E_0 -2 \omega^2)\,,
\end{equation}
result from equation (\ref{Q22}). These eigenvalues are real numbers
so that the rest energies, $E_0=\Re E_0-\frac{3i\omega}{2}$, must be
complex numbers whose imaginary parts are due to the decay produced
by the de Sitter expansion. The above results indicate that the CRs
are reducible to direct sums of UIRs of the principal series
\cite{WW}, $(p,q)$, with $p=s$ and $q(1-q)=(\Re E_0)^2+\frac{1}{4}$
\cite{Gaz}. What is new here is that we meet only one type of UIRs,
denoted by $[\Re E_0,s]$, which are completely determined by the
rest energy and the spin defined as in special relativity. In the
flat limit these UIRs tend to the corresponding Wigner ones.

Another important point is to study the covariant fields with unique
spin in the sense of the $SL(2,\Bbb C)$ symmetry. In this case,  we
must select the representations $\rho$ with unique spin, $(s,0)$ or
$(0,s)$ and, obviously, $(s,0)\oplus(0,s)$, for which we have to
replace $S^{(\rho)}_{0i}=\pm\, i S^{(\rho)}_{i}$ in equation
(\ref{Q22}) finding the remarkable identity
\begin{equation}\label{Q1Q2}
{\cal C}_2^{(s)}={\cal C}_1^{(s)}(\vec{S}^{(s)})^2-2\omega^2
(\vec{S}^{(s)})^2+\omega^2 [(\vec{S}^{(s)})^2]^2\,.
\end{equation}
Analyzing particular fields, as for example the vector field
presented in the next section, we see that this identity  holds even
for other representations $\rho$ for which the uniqueness of the
spin is guaranteed by supplemental constraints. Moreover, whether
the particles are at rest then equations (\ref{rest}) and
(\ref{rest2}) lead to the identity (\ref{Q1Q2}) for each subspace
${\cal V}_s\subset{\cal V}_{(\rho)}$ separately. This suggests that
the CRs with unique spin $s$ are equivalent with the UIRs $[\Re
E_0,s]$.

However, despite of the above results, it remains to find the
concrete form of the transformations among the CRs and the UIRs we
found here. Another problem is the relation between the rest energy
and mass since in the de Sitter case we do not have a general rule
analogous to the Poincar\' e mass condition (providing
$E_0=\pm\,m$). This impediment leads to ambiguities as we shall see
in the next section where we show that the masses defined by the
usual field equations give rise to {\em different} rest energies for
the scalar, vector and Dirac fields minimally coupled to the de
Sitter gravity.

Finally, we specify that the physical interpretation adopted here is
plausible since in the flat limit we recover the usual physical
meaning of the Poincar\' e generators. We observe that the
generators (\ref{Ji}) are independent on $\omega$ having the same
form as in the Minkowski case, $J_k^{(\rho)}=\hat J_k^{(\rho)}$. The
other generators have the limits
\begin{equation}
\lim_{\omega\to 0}H=\hat H=i\partial_t\,,\quad \lim_{\omega\to
0}(\omega R_i^{(\rho)})=\hat P_i=i\partial_i\,, \quad
\lim_{\omega\to 0}K_i^{(\rho)}=\hat K_i^{(\rho)} \,,
\end{equation}
which means that the basis $\{ H, P_i, J^{(\rho)}_i, K^{(\rho)}_i\}$
of the algebra $s(M)=sp(2,2)$ tends to the basis $\{\hat H, \hat
P_i,\hat J^{(\rho)}_i,\hat K^{(\rho)}_i\}$ of the $s(M_0)$ algebra
when $\omega\to 0$. Moreover, the Pauli-Lubanski operator
(\ref{PaLu}) is the flat limit of the five-dimensional
vector-operator (\ref{WW}) since
\begin{equation}
\lim_{\omega\to 0}W^{(\rho)}_0=\hat W^{(\rho)}_0\,,\quad
\lim_{\omega\to 0}W^{(\rho)}_i=\hat W^{(\rho)}_i\,, \quad
\lim_{\omega\to 0}W^{(\rho)}_4=0 \,.
\end{equation}
Under such circumstances the limits of our invariants read
\begin{equation}
\lim_{\omega\to 0}{\cal C}_1^{(\rho)}=\hat {\cal C}_1=\hat P^2\,,
\quad \lim_{\omega\to 0}{\cal C}_2^{(\rho)}=\hat {\cal
C}_2^{(\rho)}\,,
\end{equation}
indicating that their physical meaning may be related to the mass
and spin of the matter fields in a similar manner as in special
relativity.

\section{Covariant physical fields}

In order to investigate how the de Sitter invariants  may depend on
mass and spin we must focus only on the fields satisfying equations
that can be seen as defining the mass in each particular case
separately. These are the usual fields with unique spin, i. e, the
scalar, (vector) Proca and Dirac fields, whose Casimir operators
obey the identity (\ref{Q1Q2}).

\subsection{The massive scalar field}

The simplest example is the massive scalar field $\phi$ minimally
coupled to the de Sitter gravity. This satisfies the Klein-Gordon
equation
\begin{equation}\label{EKlein}
\frac{1}{\sqrt{g}}\,\partial_{\nu}(\sqrt{g}\,g^{\nu\alpha}\partial_\alpha
\phi) +m^2 \phi=0\,,\quad g=|\det(g_{\mu\nu})|\,,
\end{equation}
which can be put in the form ${\cal E}_{KG}\,\phi=m^2\phi$ using the
operator (\ref{EKG}). Since $\rho=(0,0)$ there are only genuine
orbital generators commuting with ${\cal E}_{KG}$ and the de Sitter
invariants ${\cal C}_1={\cal E}_{KG}\sim m^2$ and ${\cal C}_2=0$.

The rest energy of the scalar field deduced from  equation
(\ref{rest}) reads
\begin{equation}\label{EnKG}
E_0=-\frac{3i\omega}{2}\pm \omega k_s\,, \quad
k_s=\sqrt{\frac{m^2}{\omega^2}-\frac{9}{4}}\,.
\end{equation}
We remind the reader that $ik_s$ is the index of the time-dependent
Hankel functions of the plane wave solutions of the Klein-Gordon
equation \cite{Cs}.  This result can be obtained directly by solving
the mentioned equation for the rest particle with $P_i\sim 0$.

The conclusion is that the de Sitter invariants of the scalar field
minimally coupled to gravity define the UIRs $[\pm \omega k_s,0]$
and coincide to the Poincar\' e invariants of the Wigner UIRs $(\pm
m,0)$ of the flat case. For other types of coupling this property
does not hold because of the supplemental terms introduced by these
couplings \cite{BD}.

\subsection{The massive Dirac field}

The massive Dirac field $\psi$ on the de Sitter spacetime $(M,g)$ is
defined as a spinor field which transforms under gauge
transformations according to the spinor representation
$\rho_s=(\frac{1}{2},0)\oplus(0,\frac{1}{2})$ of the $SL(2,\Bbb C)$
group. The Dirac matrices (labeled by local indices)  satisfy the
identities
$\{\gamma^{\hat\alpha},\gamma^{\hat\beta}\}=2\eta^{\hat\alpha
\hat\beta}{\bf 1}_{4\times4}$ and give rise to the generators
$\rho_s(S^{\hat\mu
\hat\nu})=\frac{i}{4}\,[\gamma^{\hat\mu},\gamma^{\hat\nu}]$.  The
covariant Dirac equation, ${\cal E}_D\psi=m\psi$, of the spinor
field minimally coupled to gravity,  is governed by the Dirac
operator which has the form
\begin{equation}\label{Dirac}
{\cal E}_D=i\gamma^0\partial_{t}+ie^{-\omega t}\gamma^i\partial_i
+\frac{3i\omega}{2}\gamma^{0}\,,
\end{equation}
in the chart $\{t,\vec{x}\}$ and the gauge (\ref{gauge}). The
generators defined by equations (\ref{Ji})-(\ref{Ham}) commute with
${\cal E}_D$ offering one a large collection of operators among them
various sets of commuting operators were used for deriving different
Dirac quantum modes in four dimensions \cite{CD1,CD2}. Notice that
generalizations to higher dimensions were also proposed \cite{KP}.

The first de Sitter invariant is derived using equations (\ref{CES})
and (\ref{Dirac}) which lead to the identity
\begin{equation}\label{Q1E}
{\cal C}_1^{(\rho_s)}= {\cal E}_D^2+\frac{3}{2}\,\omega^2{\bf
1}_{4\times4} \sim m^2+\frac{3}{2}\,\omega^2\,.
\end{equation}
This result and equation (\ref{rest}) yield the rest energy of the
Dirac field,
\begin{equation}\label{EnD}
E_0=-\frac{3i\omega}{2}\pm m\,,
\end{equation}
which has a natural simple form where the decay (first) term is
added to the usual rest energy of special relativity. A similar
result can be obtained by solving the Dirac equation with vanishing
momentum.

The second invariant results from equations (\ref{Q1Q2}) and
(\ref{Q1E}) if we take into account that
$(\vec{S}^{(\rho_s)})^2=\frac{3}{4}\,{\bf 1}_{4\times4}$. Thus we
find
\begin{equation}
{\cal C}_2^{(\rho_s)}=\frac{3}{4}\,{\cal E}_D^2+\frac{3}{16}\,
\omega^2{\bf 1}_{4\times4} \sim
\frac{3}{4}\left(m^2+\frac{1}{4}\,\omega^2\right)=\omega^2
s(s+1)\nu_+\nu_-\,,
\end{equation}
where  $s=\frac{1}{2}$ is the spin and $\nu_{\pm}=\frac{1}{2}\pm
i\frac{m}{\omega}$ are the indices of the Hankel functions giving
the time modulation of the Dirac spinors of the momentum basis
\cite{CD1}.

These invariants define the UIRs $[\pm m,\frac{1}{2}]$. In the flat
limit we recover the well-known results
\begin{equation}
\lim_{\omega\to 0} {\cal C}_1^{(\rho_s)}\sim m^2\,,\quad
\lim_{\omega\to 0} {\cal C}_2^{(\rho_s)}\sim \frac{3}{4}\,m^2\,,
\end{equation}
corresponding to the  Wigner UIRs $(\pm m,\frac{1}{2})$.

\subsection{The Proca field}

The theory of the Proca (massive vector) field $A$, minimally
coupled to the gravity of the de Sitter spacetime,  is based on the
Proca equation in natural frames,
\begin{equation}\label{EQ}
\frac{1}{\sqrt{g}}\,\partial_{\nu}(\sqrt{g}\,g^{\nu\alpha}g^{\mu\beta}
F_{\alpha\beta})+m^2  A^{\mu}=0\,,
\end{equation}
where $F_{\mu \nu }=\partial_{\mu } A_{\nu }-\partial_{\nu } A_{\mu
}$ is the field strength. The Lorentz condition,
\begin{equation}\label{Lor}
\partial_{\mu}(\sqrt{g} A^{\mu})=0\,,
\end{equation}
which is mandatory for $m\not=0$, guarantees the uniqueness of the
spin $s=1$. Recently we presented the complete quantum theory of the
Proca field in the de Sitter moving chart with conformal time
\cite{Cv}. However, here we must consider the vector field in
non-holonomic frames where we can exploit the results of the
previous section.

For outlining the theory in non-holonomic frames we start with the
column matrix $A=[A^0,A^1,A^2,A^3]^T$  formed by the components
$A^{\hat\alpha}=\hat e^{\hat\alpha}_{\mu}A^{\mu}$ in the local
frames defined by the gauge (\ref{gauge}) in the chart
$\{t,\vec{x}\}$. Then, after a little calculation, we can put the
Proca equation (\ref{EQ}) in the form ${\cal E}_P A = m^2 A$
defining the Proca operator
\begin{equation}
{\cal E}_P = {\small \left[
\begin{array}{cccc}
e^{-2\omega t}\Delta&e^{-\omega
t}\partial_1(\partial_t+\omega)&e^{-\omega
t}\partial_2(\partial_t+\omega)&e^{-\omega
t}\partial_3(\partial_t+\omega)\\
-e^{-\omega t}\partial_1(\partial_t+\omega)& {\cal E}-e^{-2\omega
t}\partial_1^2&-e^{-2\omega
t}\partial_1\partial_2&-e^{-2\omega t}\partial_1\partial_3\\
-e^{-\omega t}\partial_2(\partial_t+\omega)&-e^{-2\omega
t}\partial_1\partial_2& {\cal E}-e^{-2\omega t}\partial_2^2&
-e^{-2\omega t}\partial_2\partial_3\\
-e^{-\omega t}\partial_3(\partial_t+\omega)&-e^{-2\omega
t}\partial_1\partial_3&-e^{-2\omega t}\partial_2\partial_3& {\cal
E}-e^{-2\omega t}\partial_3^2
\end{array}\right]}
\end{equation}
where $ {\cal E} = {\cal E}_{KG}-2\omega^2$. In addition, we
introduce the line matrix-operator
\begin{equation}
{\cal L}={\small \left[
\begin{array}{cccc}
\partial_t+3\,\omega&e^{-\omega t}\partial_1&e^{-\omega t}\partial_2&e^{-\omega
t}\partial_3
\end{array}\right]}
\end{equation}
which helps us to write the Lorentz condition (\ref{Lor}) simply as
${\cal L}A=0$.

The vector field $A$ is defined on the carrier space ${\cal
V}_{(\rho_v)}={\cal V}_0 \oplus {\cal V}_1$ of the usual vector
representation $\rho_v=(\frac{1}{2},\frac{1}{2})$ which is
irreducible. Therefore, the generators (\ref{Ji})-(\ref{Ham}) have
to be calculated by using the well-known matrices of the
representation $\rho_v$ \cite{W}. The spin terms of the rotation
generators act in the three-dimensional subspace ${\cal V}_1$ (with
$s=1$) while the elements of ${\cal V}_0$ behave as scalars. The
Lorentz condition which selects the spin $s=1$ without eliminating
the scalars must guarantee that the Casimir operators have the same
action on both the subspaces of  ${\cal V}_{(\rho_v)}$. Indeed, a
straightforward calculation leads to the following identities
\begin{eqnarray}
{\cal C}_1^{(\rho_v)}&=&{\cal E}_P+{\cal D L}\,,\\
{\cal C}_2^{(\rho_v)}&=&2 ({\cal C}_1^{(\rho_v)}- {\cal D
L})=2\,{\cal E}_P\,,
\end{eqnarray}
where we use the column matrix-operator
\begin{equation}
{\cal D}=\left[
\begin{array}{cccc}
-\partial_t&e^{-\omega t}\partial_1&e^{-\omega t}\partial_2&e^{-\omega t}\partial_3\\
\end{array}\right]^T\,.
\end{equation}
Hereby we observe that the Lorentz condition (${\cal L}A=0$) allows
us to drop out the term ${\cal D L}$  remaining with the eigenvalues
\begin{equation}
{\cal C}_1^{(\rho_v)}\sim m^2\,, \quad {\cal C}_2^{(\rho_v)}\sim 2
m^2\,,
\end{equation}
which {\em coincide} to those of the Wigner UIRs $(\pm m,1)$ of the
flat case.

The scalar component $A^0$ vanishes when the particle is at rest
such that the rest energy has to be calculated only for $s=1$. Using
equation (\ref{rest}) or solving directly the Proca equation we
obtain the rest energy
\begin{equation}\label{EnP}
E_0=-\frac{3i\omega}{2}\pm \omega k_v\,, \quad
k_v=\sqrt{\frac{m^2}{\omega^2}-\frac{1}{4}}\,.
\end{equation}
This means that the Proca field transforms according to the UIRs
$[\pm\omega k_v,s]$. As in the scalar case, $i k_v$ is the index of
the Hankel functions of the plane wave solutions of the Proca
equation \cite{Cv}.

\section{Concluding remarks}

We succeeded here to derive the generators of the CRs of the
$Sp(2,2)$ group induced by the finite-dimensional representations of
the $SL(2,{\Bbb C})$ group. Moreover, we expressed the Casimir
operators of these representations in closed forms focusing on the
properties of their eigenvalues that represent the principal
invariants produced by the de Sitter symmetry.

Our main goal was to find the physical meaning of these invariants
analyzing how their concrete values depend on mass, spin and the de
Sitter Hubble constant.   Our new results are independent on the
local chart and the gauge we use representing the universal values
of the de Sitter invariants summarized in the next table.
\begin{center}
\begin{tabular}{lcll}
field&spin&${\cal C}_1^{(\rho)}$&${\cal C}_2^{(\rho)}$\\
&&&\\
Klein-Gordon&0&$m^2$&0\\
Dirac &$\frac{1}{2}$&$m^2+\frac{3}{2}\,\omega^2$&
$\frac{3}{4}\,m^2+\frac{3}{16}\,\omega^2$\\
Proca &1&$m^2$&$2\, m^2$
\end{tabular}
\end{center}
The conclusion is that the invariants of the scalar and vector
fields minimally coupled to the de Sitter gravity obey the flat rule
($m^2$ and $m^2 s(s+1)$ respectively). The Dirac field behaves in a
different manner since its invariants have supplemental terms
depending on $\omega$ which vanishes in the flat limit.

The results presented here open the way to a complete theory of the
CRs on the de Sitter manifolds. The principal problem which remains
to be solved by further investigations  is to find the
transformations which assure the equivalence of our CRs with the
UIRs of given mass and spin. We hope that this could be achieved by
combining our results with those of the five-dimensional theory
\cite{Gaz}.

Another open problem is the relation between the mass and the rest
energy which does not comply with a common rule as we deduce from
equations (\ref{EnKG}), (\ref{EnD}) and (\ref{EnP}). In fact, only
the real part of the rest energy of the Dirac field respects the
usual rule ($\pm \,m$) while the scalar and vector fields have
different rest energies in the minimal coupling. This is somewhat
strange since the rest energy of the classical de Sitter geodesic
motion is just that of special relativity, $E_0=m$ (as it is shown
in Appendix). However, the group theoretical methods are not able to
enlighten this point since this relies on more general conjectures
such as the definition of the field equations and the selection of
the appropriate couplings among the covariant fields and gravity.

Finally, we note that the three examples we analyzed above are not
enough for drawing general conclusions concerning the de Sitter
invariants of the tensor fields of any rank or to speak about a
spinor anomaly. We can say only that the mass and spin of the
covariant fields defined on the de Sitter manifold have the same
meaning and play a similar role as in special relativity giving rise
to the principal invariants of the theory of free fields.

\subsection*{Appendix: Classical conserved quantities}

It is a simple exercise to integrate the geodesic equations and to
find the conserved quantities on a geodesic trajectory of the de
Sitter background. These are proportional with
$k_{(AB)\,\mu}u^{\mu}$ (where $u^{\mu}=\frac{dx^{\mu}}{ds}$) and can
be derived by using the Killing vectors (\ref{chichi1}) and
(\ref{chichi2}). We assume that in the chart $\{ t,\vec{x}\}$ the
particle of mass $m$ has the conserved momentum $\vec{p}$ of
components $p\,^i=\omega m (k_{(0i)\,\mu}-k_{(4i)\,\mu})u^{\mu}$ so
that  we can write
\begin{equation}
u^0=\frac{dt}{ds}=\sqrt{1+\frac{{p}^{\,2}}{m^2}\, e^{-2\omega t}}\,,
\qquad u^i=\frac{d{x^i}}{ds}=\frac{p\,^i}{m}\,e^{-2\omega t}\,,
\end{equation}
using the notation $p=|{\vec{p}}\,|$. Hereby we deduce the
trajectory,
\begin{equation}
{x}^i(t)={x}_0^i+\frac{p\,^i}{\omega {p}^
2}\,\left(\sqrt{m^2+{p}^{2}e^{-2\omega t_0}}-\sqrt{m^2+{p}^2
e^{-2\omega t}}\, \right)\,,
\end{equation}
of a particle passing through the point $\vec{x}_0$ at time $t_0$.

Furthermore, we calculate the other conserved quantities, i. e. the
energy,
\begin{equation}
E=\omega\, \vec{x}_0\cdot \vec{p}+\sqrt{m^2+{p}^{2}e^{-2\omega
t_0}}\,,
\end{equation}
the angular momentum $\vec{l}=\vec{x}_0\land \vec{p}$ and the
vectors
\begin{equation}
\vec{k}=-\vec{r}-\frac{1}{\omega}\,\vec{p}=\vec{x}_0
E-\vec{p}\,\chi(t_0,\vec{x}_0)\,,
\end{equation}
corresponding to the operators (\ref{Ki}) and (\ref{Ai}) for
${P}_i\to -{p^i}$ and $H\to E$ and replacing $t\to t_0$ and
$\vec{x}\to \vec{x}_0$ including in $\chi(x)$ given by equation
(\ref{z0}b). All these conserved quantities obey the identity
$E^2-\omega^2(\vec{l}\,^2+\vec{r}\,^2-\vec{k}\,^2)=m^2$ which
represents the classical version of the first invariant (\ref{C14}).
The second invariant vanishes since $W=0$ as was expected in this
spinless case.

When the  particle is at rest, staying in $\vec{x}(t)=\vec{x}_0$
with $\vec{p}=0$, then the non-vanishing conserved quantities are
the rest energy $E_0=m$  and $\vec{k}_0=-\vec{r}_0= m\vec{x}_0 $.
Thus we see that the classical rest energy on the de Sitter
spacetime is the same as in special relativity.


\begin{thebibliography}{20}

\bibitem{FF}
M. Ferraris and M. Francaviglia, in {\em Mechanics, Analysis and Geometry: 200
Years after Lagrange} Editor: M. Francaviglia (Elsevier Sci. Pub. B. V., 1991).

\bibitem{FFF}
L. Fatibene, M. Ferraris and M. Francaviglia, {\em J. Math. Phys.}
{\bf 34}, 1644 (1994); L. Fatibene, M. Ferraris  M. Francaviglia and
M. Godina, {\em Gen. Relat. and Grav.} {\bf 30}, 1371 (1998).

\bibitem{WALD}
R. M. Wald,   {\it General Relativity} (Univ. of Chicago Press:
Chicago and London 1984).

\bibitem{SW}
S. Weinberg, {\it Gravitation and Cosmology: Principles and
Applications of the General Theory of Relativity}  (Wiley, New York,
1972).

\bibitem{BD}
N. D. Birrel and P. C. W. Davies,  {\em Quantum Fields in Curved
Space} (Cambridge University Press, Cambridge 1982).

\bibitem{Gaz}
J.-P. Gazeau and M.V. Takook, {\em J. Math. Phys.} {\bf 41},  4920
(2000);P. Bartesaghi, J.-P. Gazeau, U. Moschella and M. V. Takook,
{\em Class. Quantum. Grav.} {\bf 18}, 4373 (2001); T. Garidi, J.-P.
Gazeau and M. Takook, {\em J.Math.Phys.} {\bf 44},  3838 (2003);
J.-P. Gazeau and M. Lachièze-Rey, {\tt arXiv:0802.3441}.

\bibitem{Wood}
B. Allen and T. Jacobson, {\em Commun. Math. Phys.} {\bf 103}, 669
(1986): N. C. Tsamis and R. P. Woodard, {\em J. Math. Phys.} {\bf
48}, 042306 (2007), {\tt gr-qc/0608069}; O. Bertolami and D. F.
Mota, {\em Phys. Lett. B} {\bf 444}, 96 (1999), {\tt gr-qc/9811087};
T. Prokopek, O. T\" ornkvist and R. P. Woodard, {\em Phys Rev.
Lett.} {\bf 89}, 101301 (2002); T. Prokopec, N. C. Tsamis and R. P.
Woodard,  {\em Phys. Rev. D} {\bf 79}, 043423 (2008).


\bibitem{SG}
H. B. Lawson Jr. and M.-L. Michaelson, {\em Spin Geometry}
(Princeton Univ. Press. Princeton, 1989).

\bibitem{ES}
I. I. Cot\u aescu,  {\em J. Phys. A: Math. Gen.} {\bf 33}, 9177
(2000).

\bibitem{CML}
B. Carter  and R. G. McLenaghan,  {\em Phys. Rev. D} {\bf 19}, 1093
(1979).

\bibitem{EPL}
I. I. Cot\u aescu, {\em Europhys. Lett.} {\bf 86}, 20003 (2009).


\bibitem{Cs}
I. I. Cot\u aescu, C. Crucean and A. Pop, {\em Int. J. Mod. Phys. A}
{\bf 23}, 2463  (2008).


\bibitem{Cv}
I. I. Cot\u aescu, {\em Gen. Relat. and Grav.} {\bf 42}, 861 (2010).


\bibitem{Max}
I. I. Cot\u aescu and C. Crucean, to appear in {\em Progr. Theor.
Phys.} {\bf 124} (2010).

\bibitem{CD1}
I. I. Cot\u aescu, {\em Phys. Rev. D}  {\bf 64}, 084008 (2002).


\bibitem{CD2}
I. I. Cot\u aescu, R. Racoceanu and C. Crucean, {\em Mod. Phys.
Lett. A} {\bf 21}, 1313 (2006); I. I. Cot\u aescu  and C. Crucean,
{\em Int. J. Mod. Phys. A}   {\bf 23}, 3707 (2008).

\bibitem{WW}
J. Dixmier, {\em Bull. Soc. Math. France} {\bf 89}, 9 (1961); B.
Takahashi, {\em  Bull. Soc. Math. France} {\bf 91}, 289 (1963).


\bibitem{Wigner}
G. Mackey, {\em Ann. Math.} {\bf 44}, 101 (1942)

\bibitem{W}
W.-K. Tung,  {\em Group Theory in Physics}  (World Sci.,
Philadelphia, 1984).

\bibitem{Geroch}
R. Geroch, {\em J. Math. Phys.} {\bf 9}, 1739 (1968).

\bibitem{Kos}
Y. Kosmann, {\em Comptes Rendus Acad. Sc. Paris, serie A} {\bf 264},
344 (1967); {\em id.} {\bf 262}, 289 (1966); {\em id.} {\bf 262},
394 (1966); Y. Kosmann, {\em Ann. di Matematica Pura et Appl.} {\bf
91} (1972).



\bibitem{Witt}
E. Witten, {\tt hep-th/0106109}.




\bibitem{KP} J. F. Koksma and T.
Protopek, {\em Class. Quantum Grav.} {\bf 26}, 125003 (2009).






\end{thebibliography}
\end{document}